\documentclass[11pt]{amsart}
\usepackage[varg]{txfonts} 
\usepackage{amsmath}

\numberwithin{equation}{section}
\newtheorem{theorem}{Theorem}[section]

\newtheorem{corollary}[theorem]{Corollary}

\newcommand{\tens}{\otimes}

\newcommand{\extd}{{\rm d}}

\newcommand{\lo}{{[\![}}
\newcommand{\lc}{{]\!]}}
\newcommand{\inter}{{\mathfrak{i}}}

\begin{document}
\title{Emergence of Riemannian geometry and the massive graviton}



\author{Shahn Majid}

\address{School of Mathematical Sciences\\ Queen Mary University of London\\ 327 Mile End Rd, London E1 4NS }

\email{s.majid@qmul.ac.uk}


\begin{abstract}
  We overview a new mechanism\cite{Ma:rec} whereby classical Riemannian geometry emerges out of the differential structure on quantum spacetime,
 as extension data for the classical algebra of differential forms. Outcomes for physics include a new formula for the standard Levi-Civita connection, 
  a new point of view of the cosmological constant as a very small mass for the graviton of around $10^{-33}$ev, and a weakening of metric 
  compatibility in the presence of torsion. The same mechanism also provides a new construction for quantum bimodule connections
 on quantum spacetimes and a new approach to the quantum Ricci tensor. \end{abstract}
\maketitle
\section{Introduction}
\label{intro}
It has been argued and is widely accepted that somehow classical gravity should emerge from an as yet unknown theory
of quantum gravity. The latter is expected to be some kind of algebra or combinatorics much different from the smooth
continuum, and geometric and matter fields are meant to `emerge' form the quantum gravity observables. How exactly this
is meant to happen is unclear but one now widely accepted view, which I have been pursing since the 1980s \cite{Ma:pla}, 
is that an intermediate stage should be some kind of noncommutative or `quantum' geometry. Thus, 
\begin{equation}\label{limit}{\rm Classical\ GR}\quad \Leftarrow \quad {\rm Quantum\ Geometry}\quad\Leftarrow\quad{\rm Quantum\ Gravity} \end{equation}
which in practice amounts to:

\begin{quote}{\bf Quantum spacetime hypothesis}. Spacetime is not a continuum but more effectively described by a
noncommutative coordinate algebra where $x,y,z,t$ do not fully commute.
\end{quote}

Since points in quantum spacetime cannot be simultaneously measured it means that the continuum assumption is dropped
-- spacetime becomes fuzzy below the Planck scale $\lambda$. But there are also less obvious predictions, among them
 the prediction from models of
flat quantum spacetime that the speed of light is energy-dependent\cite{AmeMa:wav}, in principle testable 
even if the noncommutativity parameter $\lambda\sim 10^{-44}$s (the Planck time).  More recent is our
prediction \cite{Ma:alm} that the classically infinite time dilation factor at a black hole
event horizon is in fact finite due to quantum spacetime effects. On the more speculative side, reported last year in \cite{Ma:icnfp1}, 
in the quantum mechanics limit of a Klein-Gordon
particle on quantum spacetime one finds that the particle's effective gravitational and inertial mass differ as the
Klein-Gordon mass parameter approaches the Planck mass. The inertial mass levels off while the gravitational 
mass decays to zero above the Planck mass.  There have also been conceptual insights from the quantum spacetime
hypothesis, among them the emergence in some models of time and of the Laplacian as its conjugate `momentum' out of the differential structure, see \cite{Ma:alm}. More recently we found the emergence in some models of curvature (gravity) as similarly being forced by the algebraic constraints of non-commutative geometry not-directly visible in the classical limit\cite{BegMa3}. The picture that emerges
is one where the quantum spacetime
hypothesis has predictive and explanatory power in the form of restrictions which are inexplicable classically but which exist classically as a remnant of the quantum geometry restriction.

In this paper we overview and discuss the physical implications of new work\cite{Ma:rec} that  goes further on the conceptual side:

\begin{quote}{\bf Proposition} The quantum spacetime hypothesis provides a mechanism for the 
emergence of Riemannian structure -- a metric and its Levi-Civita connection -- out of nothing but  the Leibniz rule.
\end{quote}

 We should be clear that this is a radical statement. That is because conventional thinking has different layers of structure, which we present as a kind of dictionary.

\bigskip
\begin{tabular}{ll}
\hline
Geometry & Algebra\\\hline
topological space  & coordinate algebra   $A$\\
differentiable manifold& differential graded algebra $\Omega(A)$ \\
Riemannian structure etc & noncommutative metric, connections etc \\\hline
\end{tabular}
\bigskip

\noindent
On the right, the algebra need not be commutative (which is then  noncommutative geometry or in our case quantum spacetime). What we 
show is that the second line {\em on the right} -- the concept of differential structure -- implies the third line {\em on the left} -- classical Riemannian geometry. 

Specifically, we are going to consider unwinding (\ref{limit}) by fixing a classical manifold and its differential structure expressed as its algebra 
$\Omega(M)$ of differential forms (and exterior differential $\extd$ obeying the graded-Leibniz rule) and then asking {\em what are all the quantum differential algebras $\widetilde\Omega(M)$ that could
project onto $\Omega(M)$ in the classical limit?}  We pose this as a central extension problem and just as group central 
extensions are controlled by cocycle data, we find \cite{Ma:rec} that extensions are described by a kind of cocycle. So this is what Riemannian
geometry {\em is}, a cocycle for quantising the classical differential structure, or put another way a remnant of the quantum differential structure. 

This new way of thinking about Riemannian geometry as emerging out of differential algebra also suggests an interpretation
of the cosmological constant or `dark energy density' as mass of the gravitational field. This is an alternative point of view in massive gravity but which tends to lead there to physically unrealistic theories\cite{massivegrav}. But our version of it is not a new theory and therefore 
is physical. It is just that we have a new mathematical point of view, see Section~\ref{graviton}, whereby the Einstein tensor appears as a non-linear wave operator $-{1\over 2}(\Delta+S)$ on the  metric $g$.  

Secondly we will see that the cocycles that arise actually correspond to a slightly larger class
of structures where the connection need not be metric-compatible in the usual way. Rather, its metric-compatibility is controlled by the torsion through
an equation which in tensor calculus notation (and indices on the torsion $T^{\mu\nu}{}_{\rho}$ raised and lowered via the metric $g$),  is
\[ T_\mu{}^{\nu\rho}+T_\mu{}^{\rho\nu}=g^{\nu\rho}{}_{;\mu}. \]
Here the semicolon denotes covariant derivative.  The Riemannian case where $T=0$ corresponds to cocycles which are cohomologous to ones where the $\extd$ is not deformed. 

Another observation will be that it is not particularly natural from the algebraic point of view for the metric that emerges to be non-degenerate, which could be useful. More radically:

\begin{quote}{\bf Proposition} The metric is really part of a covariant derivative $\nabla_\omega$ acting on all degrees of the exterior algebra
$\Omega(M)$, namely the part acting on degree 0.  \end{quote}

The covariant derivative, metric and interior product will be further unified into a single universal covariant derivative acting on all degrees and along all degrees of $\omega$. Our approach to Riemannian geometry also links up with  BRST quantisation\cite{BV} in that there is a natural Batalin-Vilkovisky (BV) algebra structure on the algebra of differential forms. 
Essentially, we find that a BV algebra equipped with a differential $\extd$ and where the BRST operator is symmetric in a certain sense (which we will explain), more or less is a Riemannian manifold. This
seems to me a remarkable confluence of a structure in quantum field theory (for the quantisation of ghosts) with gravity.

Finally, while lessons for classical GR are our main concern, our new way of thinking about Riemannian geometry also applies in the quantum case where $\Omega(M)$ is replaced by
some general differential graded algebra (DGA), which we denote  $\Omega(A)$, not necessarily graded-commutative. Then cocycle central extensions of such objects similarly 
induce (bimodule) quantum connections and `quantum interior products' giving a new approach to the latter.  This new approach provides for the first time a somewhat
general definition of the quantum Ricci tensor, see Section~\ref{noncomext}.

\section{Reconstructing a  Riemannian geometry from its divergence operator}\label{secrec}

Our claims depend on the following results taken from our recent paper \cite{Ma:rec}. Let $M$ be a smooth manifold and
let $(\Omega(M),\extd)$ be the exterior algebra of differential forms. Here $\extd$ increases degree by 1 and obeys the graded Leibniz rule and $\extd^2=0$. 

We define a possibly-degenerate metric as a symmetric tensor viewed as an `inner product' $(\ ,\ )$ on 1-forms. We define 
 a covariant derivative $\nabla_\omega:\Omega^1(M)\to \Omega^1(M)$ along 1-forms $\omega$  by the property
 \[ \nabla_\omega(a\eta)=(\omega,\extd a)\eta+a\nabla_\omega\eta,\quad\forall a\in C^\infty(M),\ \eta\in\Omega^1(M).\]
 Similarly we will define interior product by a 1-form $\eta\in \Omega^1(M)$ to  be $\inter_\eta(\omega)=(\omega,\eta)$ if $\omega$ is a 1-form and then extended by the graded-Leibniz rule. When $(\ ,\ )$ is nondegenerate then these are just covariant derivative
and interior product along the vector field $\omega^*=(\omega,\ )$. There is a notion of metric-compatibility, torsion and curvature in the generalised case\cite{Ma:rec}  reducing to the usual  notions in the nondegenerate case. This generalisation is similar to the generalisation from symplectic to Poisson geometry and could be relevant to physics in some degenerate situations.

\begin{theorem}\label{main}\cite{Ma:rec} Let $\delta:\Omega(M)\to \Omega(M)$ be a linear map that lowers degree by 1, has $\delta^2=0$ and  obeys the 6-term relation
\[ \delta(a\omega\eta)-(\delta(a\omega))\eta-a\delta(\omega\eta)-(-1)^{|\omega|}\omega \delta(a\eta)+a(\delta\omega)\eta+(-1)^{|\omega|}a\omega\delta\eta=0\]
for all  $a\in C^\infty(M),\ \omega,\eta\in\Omega(M)$, 
and the `symmetry condition'
\[ \delta(a\extd b)-a\delta\extd b=\delta(b\extd a)-b\delta\extd a,\quad\forall a,b\in C^\infty(M).\]
Then
\[ (\omega,\extd a)=\delta(a\omega)-a\delta\omega\]
\[ \nabla_{\omega}\eta={1\over 2}\left(\delta(\omega\eta)-(\delta\omega)\eta+\omega\delta\eta+\inter_{\omega}\extd\eta+\inter_{\eta}\extd\omega+\extd(\omega,\eta)\right)\]
for all $a\in C^\infty(M),\ \omega,\eta\in \Omega^1(M)$, 
define a possibly-degenerate metric and metric-compatible torsion-free  covariant derivative.
\end{theorem}
The proof is in \cite{Ma:rec} but to explain a little, specialising the 6-term identity to $\eta=b$ a function we find 
\[ \delta(a\omega b)-a(\delta(\omega b))=(\delta(a\omega)-a\delta \omega)b,\quad \forall a,b\in C^\infty(M),\ \omega\in\Omega^1(M)\]
which we can read both as $(\omega b)^*(a)=\omega^*(a)b$ and as $\omega^*(ab)=\omega^*(a)b+a\omega^*(b)$ where $\omega^*(a)=\delta(a\omega)-a\delta\omega$. This means that $\omega^*$ is a vector field and depends tensorialy on $\omega$. Hence it corresponds to a bivector $(\ ,\ )$ which moreover is symmetric under the symmetry condition on $\delta$. In the case where $(\ ,\ )$ is nondegenerate we obtain a Riemannian metric and a new formula for its Levi-Civita connection on $M$. In the converse direction it is shown in \cite{Ma:rec} that we can take $\delta$ to be the divergence or codifferential, so any Riemannian structure can be obtained by this theorem. 

\begin{corollary}\cite{Ma:rec} A manifold has the structure of a Riemannian manifold iff its exterior algebra has the structure of a BV algebra with $\delta$ symmetric and with the associated bilinear nondegenerate. In this case $\delta$ can be taken to be the divergence or codifferential.
\end{corollary}
Indeed, in Theorem~\ref{main} and when the metric is non-degenerate, one finds\cite{Ma:rec} the 7-term identity
\[ \delta(\omega\eta\zeta)=(\delta(\omega\eta))\zeta+(-1)^{|\omega|}\omega\delta(\eta\zeta)+(-1)^{(|\omega|-1)|\eta|}\eta\delta(\omega\zeta)\]
\[ \qquad - (\delta\omega)\eta\zeta-(-1)^{|\omega|}\omega(\delta\eta)\zeta-(-1)^{|\omega|+|\eta|}\omega\eta\delta\zeta\]
for all $\omega,\eta,\zeta\in \Omega(M)$ as an extension of the assumed 6-term identity. A BV algebra is a degree -1 operator $\delta$ (usually denoted differently) on a graded-commutative algebra with $\delta^2=0$ and this 7-term identity, so this explains one direction. In the symmetric case and where the algebra is the exterior algebra on a manifold, we apply Theorem~\ref{main}. Whenever a graded-commutative algebra has a differential $\extd$  we can think if it as the exterior algebra on a `manifold' in some algebraic sense. So this is a generic interpretation of BV algebras with differential and with $\delta$ symmetric and non-degenerate in the manner that we have stated. The theory in \cite{Ma:rec} is more general and only needs $\delta^2$ tensorial.

\subsection{Massive graviton}\label{graviton}

A second mathematical result in \cite{Ma:rec} in the setting of Theorem~\ref{main} concerns the Hodge Laplacian
\[ \Delta=\extd\delta+\delta\extd.\]
This is a degree 0 operator on $\Omega(M)$ but it is shown that it extends to tensor products in a canonical way, in particular to a canonical action
of $\Delta$ on 1-1 forms wherein lives the metric $g$ inverse to $(\ ,\ )$. 

\begin{theorem}\cite{Ma:rec} The Hodge Laplacian extended to 1-1-forms obeys
\[ -{1\over 2}\Delta g={\rm Ricci}\]
\end{theorem}

We define $S=(\ ,\ ){\rm Ricci}$ in the usual way as the Ricci scalar. Then 
\[ {\rm Einstein} = -{1\over 2}( \Delta + S) g,\quad S=-{1\over 2}(\ ,\ )\Delta g.\]
The Hodge Laplacian provides a natural wave operator with the correction to $\Delta$ attributable to the nonlinear nature of the problem. In this case Einstein's vacuum equation with cosmological $\Lambda$ term has the form
\[  ((\Delta+S) - 2\Lambda)g=0\]
where 
\[ \sqrt{2|\Lambda|}\approx 5 \times 10^{-33}{\rm ev}\]
if we put in the observed vacuum density of around $10^{-29}$g/cm${}^3$. As the observed value appears to be positive, this would seem to imply a tachyonic mass, but this could be a matter of reviewing the sign and signature conventions. In any event we provide a point of view which puts the puzzle of the cosmological constant on the same footing as the lack of explanation of other small masses in physics, such as neutrino or other elementary particle  masses, which are all far below the Planck mass. For a possible explanation of that, there are at least some situations, eg\cite{Ma:mom}, where quantum spacetime prefers masses to be zero. These then might may appear as quantum corrections when viewed classically.

\section{Extension theory for differential graded algebras}\label{extension}

Let's now see\cite{Ma:rec} how  Riemannian geometry in the form above emerges out of nothing but noncommutative differential calculus, basically the Leibniz rule. The idea is to assume only that the differential structure on classical spacetime is the projection of some quantum differential graded algebra  $(\tilde\Omega(M),\tilde\extd)$. Thus $\tilde\Omega(M)=\oplus_n\tilde\Omega^n(M)$ plays the role of differential forms of different degrees and we suppose that $\tilde\Omega^0(M)=C^\infty(M)$ as a vector space 
and that $\tilde\extd$ increases degree by 1, is a graded-derivation and $\tilde\extd^2=0$. This is a minimal requirement. We also suppose:
\[ \tilde\Omega(M) \twoheadrightarrow \Omega(M)\]
is an extension of differential graded algebras by a single additional 1-form $\theta'$ with $\theta'^2=0, \extd\theta'=0$ and $\theta'\wedge\omega=(-1)^{|\omega|}\omega\wedge\theta'$ for all $\omega\in\Omega(M)$. So we are supposing just that the noncommutative geometry has a single extra cotangent direction, setting which to zero takes us back to the classical $\Omega(M)$. Finally, we add one key requirement, namely that the product by functions from the left is the same as classical. This is a kind of `normal ordering' assumption and such an extension is called {\em cleft} in \cite{Ma:rec}. This also entails that the product of $C^\infty(M)$ itself is not deformed, only the differentials are `quantum'. So these are only mildly noncommutative extensions of the classical manifold.

Now the general form of $(\tilde\Omega(M),\tilde\extd)$ by degree counting is \cite{Ma:rec}
\[ \tilde\extd \omega=\extd\omega-{\lambda\over 2}\theta' \Delta \omega,\quad \omega\tilde\wedge\eta=\omega\wedge\eta-{\lambda \over 2}\theta'\lo \omega,\eta\lc\]
for a degree 0 map $\Delta$ and a degree -1 (i.e. degree lowering) map $\lo\ ,\ \lc$. The constant $\lambda$ is physically dimensionful but could otherwise be absorbed in the normalisation of $\theta'$. 
 Such maps $(\Delta,\lo\ ,\ \lc)$ define an extension iff $[\Delta,\extd]=0$ and the `cocycle conditions' \cite{Ma:rec}
\begin{equation}\label{c1} \lo\omega\eta,\zeta\lc+\lo\omega,\eta\lc\zeta=\lo\omega,\eta\zeta\lc+(-1)^{|\omega|}\omega\lo\eta,\zeta\lc\end{equation}
\begin{equation}\label{c2} L_\Delta(\omega,\eta)=\extd\lo\omega,\eta\lc+\lo\extd\omega,\eta\lc+(-1)^{|\omega|}\lo\omega,\extd\eta\lc \end{equation}
hold for all $\omega,\eta,\zeta\in \Omega(M)$. For any operator $B$ of degree $b$ on a graded algebra, we use  the Leibnizator
\[ L_B(\omega,\eta):=B(\omega\eta)-(B\omega)\eta-(-1)^{b|\omega|}\omega B\eta.\]
The cleft case has the further requirement  $\lo a,\ \lc=0$ for all $a\in C^\infty(M)$.   

\begin{theorem}\label{Deltaconn}\cite{Ma:rec} Associated to any cleft-extension $(\tilde\Omega(M),\tilde\extd)$ of $(\Omega(M),\extd)$ is a classical possibly-degenerate metric and covariant derivative
\[ (\omega,\extd a)={1\over 2}\lo \omega,a\lc,\quad \nabla_\omega\eta={1\over 2}\lo \omega,\eta\lc,\quad \forall a\in C^\infty(M),\ \omega,\eta\in \Omega^1(M)\]
obeying
\begin{equation*}(\omega,\extd(\eta,\zeta)) -(\nabla_\omega\eta,\zeta)-(\eta,\nabla_\omega\zeta)=T(\omega,\eta)(\zeta)+T(\omega,\zeta)(\eta),\quad\forall\omega,\eta,\zeta\in \Omega^1(M).\end{equation*}
where $T$ is the torsion tensor. 
\end{theorem}
The details are in \cite{Ma:rec} but the idea is that in the cleft case the cocycle conditions (\ref{c1})-(\ref{c2}) specialise nicely when one of the arguments is of degree 0. From (\ref{c1}) one finds that  $\lo ,a\lc$ is a degree -1 graded-derivation so behaves as interior product along a vector field allowing us to identify it as the interior product by 1-forms with respect to some map $(\ ,\ )$ as stated. Then a case of (\ref{c2}),
\begin{equation}\label{l0} L_\Delta(a,\omega)=\lo \extd a,\omega\lc\end{equation}
tells us when applied to $\omega=b\in A$ that $(\ ,\ )$ is symmetric. After this, (\ref{c1}) tells us that $\nabla_\omega$ behaves as a covariant derivative. Finally, another case of (\ref{c2}), 
\begin{equation}
\label{l00}L_\Delta(\omega,a)=\extd\lo\omega,a\lc+\lo\extd\omega,a\lc+(-1)^{|\omega|}\lo\omega,\extd a\lc 
\end{equation}
and graded-commutativity tells us that 
\begin{equation}\label{connsym} \nabla_\eta\omega+\nabla_\omega\eta=\extd (\omega,\eta)+\inter_{\eta}(\extd\omega) +\inter_{\omega}(\extd\eta),\quad\forall \omega,\eta\in\Omega^1(M).\end{equation}
From this formula it is easy to see that the torsion tensor
\[T(\omega,\eta)(\zeta):= (\omega,\nabla_\eta\zeta)-(\eta,\nabla_\omega\zeta)-\inter_\omega\inter_\eta\extd\zeta,\quad\forall\omega,\eta,\zeta\in\Omega^1(M)\]
obeys the stated condition. 

This theorem says that our covariant derivative induced by a general cleft extension may not be metric-compatible but the failure of metric-compatibility is determined by the torsion. If the torsion is zero then we will have a metric and Levi-Civita connection coming out of the cleft extension. To describe this torsion free case we say that an extension is {\em flat} if it has coboundary $\Delta$ in the sense
\[ \Delta =\extd \delta+\delta\extd\]
for some linear map $\delta$  that lowers degree by 1. This means that the extension is isomorphic (in a kind of gauge equivalent sense) to one with $\Delta=0$ i.e. to one where $\extd$ need not be deformed. 

\begin{theorem}\cite{Ma:rec}\label{flatconn} Flat cleft extensions of $\Omega(M)$ such that $\delta^2=0$ and $\delta$ symmetric are equivalent to (possibly degenerate) Riemannian structures on $M$ via Theorem~\ref{main}.  \end{theorem}
The details are in \cite{Ma:rec} but the idea is to use (\ref{l0}) which computes as $\lo\extd a,b\lc=L_\Delta(a,b)=2(\delta(a\extd b)-a\delta\extd b)$ on making use of the symmetry assumption. Hence the definition of $(\ ,\ )$ reduces by this to the same as in Theorem~\ref{main}. The 6-term identity is then equivalent to a case of (\ref{c1}) as explained in \cite{Ma:rec}. The $\delta^2$ condition and the symmetry condition stated are not too essential  but do hold in the case of the standard divergence operator on a Riemannian manifold. 

So the key features of Riemannian geometry, namely a symmetric (but possibly degenerate) metric and its metric compatibility are induced by the existence of a noncommutative differential calculus extension of the flat type considered. This is also some of the mathematics underlying the construction of the noncommutative Schwarzschild black hole in \cite{Ma:alm} as explained in later sections of \cite{Ma:rec}. We also note a different context  where a non-cleft but equivalent to zero extension has been applied to understand stochastic differentials\cite{Beggs}.

\subsection{Novel features for classical Riemannian geometry}

What can we learn for physics from this new point of view of Riemannian geometry?

\subsubsection{Torsion connections are not naturally metric compatible} We have seen that the metric and covariant derivative pairs that most naturally arise from the noncommutatuve differential algebra are a bit beyond Levi-Civita with torsion and metric-compatibility mutually constrained by the formula in Theorem~\ref{Deltaconn}.  This is a novel prediction for the manner in which metric-compatibility could break down in generalised GR and one could look for this effect. 

\subsubsection{Unification of metric and covariant derivative}  On the conceptual side, we learn from Theorem~\ref{Deltaconn} that the metric in the form of interior product and the covariant derivative are really unified into a single `universal covariant derivative' 
\[ \nabla_\omega\eta={1\over 2}\lo\omega,\eta\lc,\quad \forall \omega,\eta\in \Omega(M)\]
where we act on all degrees and along all degrees. If $\omega\in \Omega^1$ then $\nabla_\omega$ is a natural covariant derivative acting on all degrees,  and in particular 
\[ \nabla_\omega a=(\omega,\extd a),\quad\forall\omega\in \Omega^1(M),\ a\in C^\infty(M)\]
 encodes the metric (and its extension to all degrees of $\omega$ encodes the interior product $\nabla_\omega a=\inter_{\extd a}\omega$).  The curvature and torsion vanish automatically on degree 0.

The main consequence for classical geometry is that it is not essential to ask for the metric to be nondegenerate (after all, we do not usually expect this for covariant derivatives) and indeed one has basic features  even in the degenerate case provided we work with covariant derivatives $\nabla_\omega$ along forms $\omega$ and not along vector fields as usually done.

\subsubsection{Covariant derivative along 2-forms}  

We also have $\nabla_\omega$ acting along all degrees of $\omega$. Here $\nabla_a=0$ along a function while in general (\ref{c1}), which now reads as 
\begin{equation}\label{connuni}\nabla_\omega(\eta\zeta)=(\nabla_\omega\eta)\zeta-(-1)^{|\omega|}\omega\nabla_\eta\zeta+\nabla_{\omega\eta}\zeta,\quad\forall \omega,\eta,\zeta\in \Omega(M),\end{equation}
tells us that $\nabla$ along 2-forms or higher are determined by the covariant derivative along 1-forms and also that $\nabla_\omega$ of any degree behave like covariant derivatives but with respect to $\inter_{\extd a}$. In some cases, notably when $(\ ,\ )$ is nondegenerate and the extension is flat (so that we are in the Levi-Civita case) one can show that $\nabla_\omega$ is a degree $|\omega|-1$ graded-derivation and hence find for example along 2-forms that
\begin{equation*} \nabla_{\omega\eta}=\eta\nabla_\omega-\omega\nabla_\eta ,\quad\forall \omega,\eta\in \Omega^1(M),\end{equation*}
which operator looks somewhat like an infinitesimal `angular momentum' operator in the sense
\[ \nabla_{\extd x^i\extd x^j}=\extd x^j\nabla^i-\extd x^i\nabla^j;\quad \nabla^i:=\nabla_{\extd x^i}.\]
Also, the 2-form operator $\nabla_{\omega}$, $\omega\in \Omega^2(M)$, has divergence $\delta\nabla_{\omega}$ which is essentially the curvature operator associated to $\omega$. This  could be relevant to a geometrical understanding of stress-energy.

\subsubsection{Wave equation} The fourth important lesson at the moment is that the Riemannian geometry comes associated with a Laplacian $\Delta:\Omega\to \Omega$ as a byproduct of the quantum differential calculus. It obeys (\ref{c2}) as
\begin{equation}\label{Deltauni} [\Delta,\extd]=0,\quad {1\over 2}L_\Delta(\omega,\eta)=(\extd\nabla_\omega+(-1)^{|\omega|}\nabla_\omega\extd) \eta+\nabla_{\extd\omega}\eta,\quad\forall \omega,\eta,\zeta\in \Omega(M)\end{equation}
and `emerges' at the same time as the Riemannian geometry, perhaps explaining the ubiquitous role of the wave equation in physics. Thus a zero mass eigenfunction of the wave operator, i.e. $\omega\in \Omega(M)$ such that $\Delta \omega=0$, is the same thing as an element of the exterior algebra such that
\[\tilde\extd \omega=\extd\omega,\]
in other words fields on which $\extd$ is not deformed in the noncommutative extension. {\em This is the meaning of the `wave equation' in our new point of view.} 

\section{Emergence of quantum metrics and bimodule covariant derivatives}\label{noncomext}

If the above is the right way to think about classical Riemannian geometry then it should also be the right way to think about the axioms of quantum or noncommutative Riemannian geometry where  $A$ is not necessarily given by a classical manifold, indeed it need not even be commutative. 

Recall that $A$ will play a role like a coordinate algebra. We suppose it extends to a differential graded algebra $\Omega(A)=\oplus_i\Omega^i$ playing the role of forms of different degrees and equipped with $\extd$, the `exterior derivative', obeying $\extd^2=0$ and the graded-Leibniz rule. We already met this idea when talking about the central extension of the classical exterior algebra in Section~3, but now it will be our starting point. The other basic idea from noncommutative geometry that we will need is that of a {\em bimodule}, meaning that $A$ acts from both the left and the right and these actions commute. For example, each $\Omega^i$ is a bimodule meaning $(a\omega)b=a(\omega b)$ for all $a,b\in A, \omega\in\Omega^i$, at least in a conventional associative setting. Now look at 1-1 tensors like the metric $g\in \Omega^1\tens_A\Omega^1$. The subscript $A$ in the tensor product means that we can associate any `functional dependence' either with the left copy or the right copy of $\Omega^1$, in the sense $\omega\tens_A a\eta=\omega a\tens_A\eta$ for all $a\in A$. Finally, when we talk about tensorial maps in geometry we mean that they commute with multiplication by functions on the space, which we can now do on either side when we have a bimodule. Such {\em bimodule maps} which commute with multiplication from either side are the right notion then of `tensoriality' in noncommutative geometry in the approach being taken here. One could also have left-tensoriality and right-tensoriality separately. 

Now suppose some initial $(\Omega(A),\extd)$ of standard type generated by $A,\extd$ as a differential graded algebra but possibly non-graded-commutative, and we consider similar extensions 
\[ \tilde\Omega(A) \twoheadrightarrow \Omega(A)\]
where $(\tilde\Omega(A),\tilde\extd)$ has an extra generator $\theta'$ as before. We have the same notions as before and a cleft extension is governed by $\lo a,\omega\lc=0$ and obeys (\ref{c1})-(\ref{c2}) as before, just this time with $a\in A$ and $\omega,\eta\in \Omega$.  We then  follow the line of Theorem~\ref{Deltaconn} by setting
\[ j_\omega( a\extd b)={1\over 2} \lo\omega a, b\lc,\quad\forall \omega\in \Omega,\ a,b\in A\]
which we suppose gives a well-defined map $j:\Omega\tens_A\Omega^1\to \Omega$ as a kind of `regularity' assumption. In this case  (\ref{c1})  means that   the map $j$ or `interior product' is a bimodule map, i.e `tensorial' as explained above. In our approach this map $j$ extends the role of metric, where the latter now appears as the restriction of the interior product 
\[ (\omega,\zeta)=j_\omega(\zeta),\quad \forall \omega,\zeta\in\Omega^1.\]
Here $j_\omega(\zeta)$ reduces to $\inter_\zeta(\omega)$ in the classical case of Section~2, i.e. we adjusted the notation better to fit the noncommutative case. 

Then on 1-forms, different instances of (\ref{c1}) now tell us that 
\[ \nabla_\omega\eta={1\over 2}\lo\omega,\eta\lc,\quad\forall \omega,\eta\in \Omega\]
is an example of a `bimodule covariant derivative' in the sense
\begin{equation}\label{nablagen} \nabla_{a\omega}=a\nabla_\omega,\quad \nabla_\omega(a\eta)=j_\omega(\extd a)\eta+ \nabla_{\omega a}\eta\end{equation}
\begin{equation}\label{sigma} \nabla_\omega(\eta a)=(\nabla_\omega\eta)a+\sigma_\omega(\eta\tens_A\extd a);\quad \sigma_\omega(\eta\tens_A\zeta)=j_{\omega\eta}(\zeta)-(-1)^{|\omega|}\omega j_\eta(\zeta) \end{equation}
for all $a\in A, \omega,\eta\in\Omega$, where $\sigma:\Omega\tens_A\Omega\tens_A\Omega^1\to \Omega$ is a bimodule map. The idea of a bimodule connection is somewhat accepted now in noncommutative geometry cf\cite{MDV,BegMa2} and it is remarkable that we get one of these out of a cleft extension, albeit with a very special form of the generalised braiding $\sigma$ and expressed as a covariant derivative along forms of all degrees.

Having a bimodule covariant derivative in the case where $(\ ,\ )$ is invertible, we obtain a well-defined covariant derivative on tensor products, eg 
\[ \nabla_\omega(\eta\tens_A\zeta)=\nabla_\omega(\eta)\tens_A\zeta+ \sigma_\omega(\eta\tens_A g^1)\tens_A\nabla_{g^2}(\zeta),\quad\forall \omega,\eta,\zeta\in \Omega^1.\]
where $g=g^1\tens_A g^2$ is a short notation for the  inverse of $(\ ,\ )$. We can then define metric compatibility as $\nabla_\omega g=0$ for all $\omega\in\Omega^1$, i.e. we have good definitions  and the main ingredients for quantum Riemannian geometry, and a specific form for the covariant derivative and interior-product-metric.

As in the classical case,  a general regular cleft extension of $\Omega(A)$ will not give a torsion free or a metric compatible covariant derivative. But this time when $\Omega(A)$ is non-graded-commutative we no longer have $L_\Delta(a,b)=L_\Delta(b,a)$ and $L_\Delta(a,\omega)=L_\Delta(\omega,a)$ meaning that $(\ ,\ )$ is no-longer symmetric and  $\nabla$ no longer has symmetric part (\ref{connsym}). However, if our differential graded algebras deform the classical case then these desirable features will still be true up to the noncommutativity parameter $\lambda$. 

Moreover, to get to a quantum version of the Levi-Civita case we need according to Theorem~\ref{flatconn} to find a flat cleft extension, i.e. where $\Delta=\extd\delta+\delta\extd$ for some $\delta$ that lowers degree by 1. The main result in this section of \cite{Ma:rec} is a new construction for this that works equally well in the quantum case. 

\begin{theorem}\cite{Ma:rec}\label{perptheorem}  Let $\Omega(A)$ be a standard $DGA$ and let $\perp$ be a degree -2 `product' on it such that $\perp a=a\perp=0$ for all $a\in A$ and 
\[ (-1)^{|\eta|}(\omega\eta)\perp\zeta+(\omega\perp\eta)\zeta=\omega\perp(\eta\zeta)+(-1)^{|\omega|+|\eta|}\omega(\eta\perp\zeta),\quad\forall \omega,\eta,\zeta\in \Omega,\]
and $\delta$ a regular degree -1 map on $\Omega$ such that 
\[\delta(a\omega)-a\delta\omega=\extd a\perp\omega,\quad \forall a\in A,\ \omega\in \Omega.\]
Then there is a regular flat cleft extension with Laplacian and associated quantum inner product and quantum bimodule covariant derivative
\[ \Delta=\extd\delta+\delta\extd,\quad j_\omega(\extd a)={1\over 2}(\delta(\omega a)-(\delta\omega)a+\omega\perp \extd a)\]
\[  \nabla_\omega\eta={1\over 2}\left(L_\delta(\omega,\eta)+\omega\perp\extd\eta-(-1)^{|\omega|}\extd\omega\perp\eta-(-1)^{|\omega|}\extd(\omega\perp\eta)\right)\]
 The cocycle is given by $\Delta$ and $\lo\omega,\eta\lc=2\nabla_\omega\eta$ for all $\omega,\eta\in \Omega$. 
\end{theorem} Details are in \cite{Ma:rec} but the idea is to solve first with $\Delta=0$ as the flat case is equivalent to this, but not necessarily cleft. This comes down to finding the map $\perp$. We then `gauge transform' to an equivalent cleft extension by adding the coboundary of a suitable $\delta$. We require it to be regular in the sense that $j$ is well-defined by the formula given. We also have $\sigma$ from (\ref{sigma}). 

What the theorem amounts to is a quantum version of the construction that associates to a metric its Levi-Civita connection, something which has largely been missing in noncommutative geometry. We  have not supposed an analogue of $\delta^2=0$ (nor the weaker $\delta^2$ tensorial) nor symmetry of $\delta$, so we will not necessarily land on a deformation of the classical Levi-Civita connection exactly until we add such restrictions. 

\subsection{Novel features in the noncommutative case}

What new lessons can we learn for noncommutative geometry from this?  

\subsubsection{New approach to the interior product}

The key new lesson in the quantum case is that we don't need just the metric inner product on degree 1 but $\perp$ on all degrees subject to the novel 4-term relation shown in Theorem~\ref{perptheorem}. In the classical case $\perp$ is automatically obtained by extending as a derivation from the metric in degree 1 and that is just what we don't have for a general $\Omega(A)$, but we've found a replacement. 

More generally, inner products in noncommutative geometry have not been fully understood and we have seen a new approach to them in which they arise as a bimodule map $j: \Omega\tens_A\Omega^1\to \Omega$
from our universal covariant derivative acting in degree 0. From this point of view, the technical regularity assumption for a regular cleft assumption amounts to the existence of a bimodule covariant derivative. Thus, suppose
\[ \nabla_\omega(ab)=j_\omega(\extd a)b+ \nabla_{\omega a} b\]
for some map $j$. Setting $b=1$ tells us that indeed $\nabla_\omega (a)=j_\omega(\extd a)$. Also assume the bimodule covariant derivative property in degree 0,
\[ \nabla_\omega(ab)=(\nabla_\omega a)b+\sigma_\omega(a\tens_A\extd b)\]
which, comparing with the left covariant derivative property, amounts to $\sigma_\omega(a\tens_A\extd b)=j_{\omega a}(\extd b)$. Then $\sigma$ being well-defined and a bimodule map forces us to take $j$ a bimodule map obeying $j_{\omega}(a\extd b)=j_{\omega a}(\extd b)$. So our regularity assumption is really just the logic of requiring a bimodule covariant derivative acting in degree 0. 

Also the interior product in our case is part of our universal bimodule covariant derivative $\nabla_\omega\eta$ for all degrees of $\omega,\eta$. Perhaps this is also a lesson for connections. After all, the interior product $j_\omega(\zeta)$ and its cousin $\omega\perp\zeta$ are needed not only for degree 1 but for all degrees of $\omega$ so why should not $\nabla_\omega$ also extend? In nice cases the covariant derivative along higher degrees can be built up from lower ones, notably in the nice case where $\nabla_\omega$ for $\omega$ of degree 1 is a braided-derivation with respect to $\sigma$, see\cite{Ma:rec} for details. But just as for interior products, we can't always assume this and have to work more generally with the generalised Leibniz rule (\ref{connuni}) that mixes covariant derivatives along different degrees. The higher form covariant derivative  also features in the properties (\ref{Deltauni}) of $\Delta$. 

\subsubsection{Torsion} At least in the case of a right-inverse for $(\ ,\ )$, one can use a standard definition of Riemann curvature and torsion of a connection as operations (otherwise we only have these evaluated against 1-forms as in Section~\ref{extension}). The curvature vanishes acting on degree 0. Torsion is defined on all degrees as a degree increasing map $\Omega\to \Omega$,
\[ T:=g^1\nabla_{g^2}-\extd\]
and also vanishes acting on degree 0. We want $T$ to be `tensorial' and as we explained, this means that we want it to be a bimodule map (one says that the connection is `torsion compatible'\cite{BegMa2}). It is already defined to be left-tensorial in the sense of commuting with multiplication from the left but for a bimodule connection we can ask for it to be fully tensorial, which comes down to \begin{equation}\label{torcomp} \wedge(\sigma(\omega\tens_A\zeta))=(-1)^{|\omega|}\omega\zeta\quad\forall \omega\in \Omega,\quad\zeta\in \Omega^1 \end{equation}
where $\sigma=g^1\tens_A\sigma_{g^2}$ is the usual generalised braiding or `flip map' in this context. Putting in our specific form of $\sigma$, torsion compatibility for covariant derivatives that come from cleft extensions amounts to
\begin{equation}\label{torcompj} g^1 g^2j_\omega(\zeta)+ g^1j_{g^2\omega}(\zeta)=(-1)^{|\omega|}\omega\zeta \quad\forall \omega\in \Omega,\quad\zeta\in \Omega^1.\end{equation}
 In nice cases where $\nabla_\omega$ is a braided-derivation one has that $T$ is then a graded-derivation, see \cite{Ma:rec}. 
   
 \subsubsection{Ricci tensor} This is not fully understood in noncommutative geometry although there are several reasonable-looking examples on a case by case basis. However, for any left covariant derivative one has a well-defined `half curvature' operator
 \[ \rho(\omega\tens_A\eta)=\nabla_\omega\nabla_{\eta}-\nabla_{\nabla_\omega \eta},\quad\forall \omega,\eta\in \Omega^1.\]
 As a result one has in the case of $(\ ,\ )$ invertible a well-defined Laplace-Beltrami operator\cite{Ma:rec} 
 \[\Delta_{LB}:=\rho(g)= \nabla_{g^1}\nabla_{g^2}-\nabla_{\nabla_{g^1}g^2}\]
and an interesting result in \cite{Ma:rec} is that in the case of a bimodule covariant derivative,
\[ L_{\Delta_{LB}}(a,\omega)=\nabla_{\extd a+\sigma_{g^1}(g^2\tens\extd a)}\omega,\quad\forall a\in A,\ \omega\in\Omega.\]
Hence if the generalised braiding is such that 
\begin{equation}\label{sigmacompat}\sigma_{g^1}(g^2\tens\zeta)=\zeta,\quad \forall \zeta\in \Omega^1,\end{equation} we will have that 
$L_{\Delta_{LB}}(a,\omega)=L_{\Delta}(a,\omega)$ if the Hodge laplacian has the usual Leibnizator $2\nabla_{\extd a}\omega$ is in our case from (\ref{Deltauni}). Then 
 $\nabla_{LB}-\Delta$ commutes with left-multiplication by $A$, i.e. is left-tensorial. For this class of bimodule covariant derivatives we can therefore define\cite{Ma:rec} 
\[ {\rm Ricci}:=g^1\tens_A (\Delta_{LB}-\Delta)(g^2)\]
which is now well-defined as it depends only on $g^1\tens_A g^2$. This short-circuits the construction of ${\rm Ricci}$ from the quantum Riemannian curvature, which has so far proven problematic, but necessarily gives the same answer in the classical case. Its further properties remain to be studied but at least we have a definition of Ricci for a class of quantum bimodule covariant derivatives. 

For our covariant derivatives that come from cleft extensions one finds that (\ref{sigmacompat}) holds whenever the metric is quantum symmetric in the sense that it is killed by the wedge product, i.e. $g^1g^2=0$, which is something we usually impose in any case in noncommutative models, so this is a very reasonable new approach. 
  
 \subsubsection{Inner calculi and a quantum example} A typical phenomenon in noncommutative geometry which is, however, impossible in classical geometry, is that there can exist a 1-form $\theta\in\Omega^1$ such that $\extd \omega=\theta \omega-(-1)^{|\omega|}\omega\theta$ for all $\omega\in\Omega$. Such a calculus is called {\em inner}. This is not possible classically because classical forms graded-commute but is quite typical for a sufficiently noncommutative geometry for algebraic reasons. In this inner case one can show that $\delta$ in Theorem~\ref{perptheorem} exists once we have solved for $\perp$, indeed  in this case\cite{Ma:rec}
 \[ \delta=\theta\perp,\quad \Delta =2\nabla_\theta-\theta^2\perp,\quad j_\omega(\zeta)={1\over 2}\omega\perp\zeta\]\[
  \nabla_\omega\eta=-{1\over 2}L_{\perp\theta}(\omega,\eta),\quad \sigma_\omega(\eta\tens_A\zeta)={1\over 2}\left((\omega\eta)\perp\zeta-(-1)^{|\omega|}\omega(\eta\perp\zeta)\right)\]
for all $\omega,\eta\in\Omega,\ \zeta\in\Omega^1$. In particular, $(\ ,\ )={1\over 2}\perp$ restricted to degree 1. Also on the coordinate algebra, $\Delta a=2(\theta,\extd a)=\theta\perp\extd a$ in keeping with the analysis in \cite{Ma:gra}, which is now extended to all degrees of forms.  
 
 To illustrate, we summarise quantum geometry on 2 points $\{x,y\}$ from  \cite{Ma:rec}.  Let $A$ be the commutative algebra of functions on this set. Like any unital algebra this has a maximally noncommutative `universal' exterior algebra $\Omega$. In our case it is inner and generated by $A$ and a single 1-form $\theta$, so in degree $n$ the $n$-form $\theta^n$ is a basis.  If $f\in A$, let $\bar f(x)=f(y)$, $\bar f(y)=f(x)$. Then the structure of $\Omega$ is
 \[ \theta f= \bar f \theta,\quad\extd f=(\bar f-f)\theta,\quad \extd \theta^n= (1-(-1)^{n})\theta^{n+1}.\]
 On this one can solve the 4-term relation for $\perp$ as\cite{Ma:rec}
 \[ \theta^m\perp \theta^n=2 (-1)^{m+1}mn \theta^{m+n-2}\]
where 2 fixes a particular normalisation. This induces by Theorem~\ref{perptheorem} the coderivation, interior product, bimodule covariant derivative, inverted metric and  Laplacian\cite{Ma:rec}, \[ \delta (f\theta^n)=2\bar f n \theta^{n-1},\quad j_{\theta^m}(\theta)=(-1)^{m+1}m \theta^{m-1}\]
\[  \nabla_\theta (f \theta^n)=(f-(-1)^n\bar f)\theta^n,\quad \sigma_\theta(\theta^n\tens \theta)=(-1)^n\theta^n\]
\[ g=\theta\tens \theta,\quad\Delta\omega=2(\nabla_\theta\omega+2 |\omega|\omega),\quad\forall \omega\in\Omega.\]
 One finds\cite{Ma:rec} that {\em this quantum covariant derivative is torsion free and metric-compatible}.  Thus we obtain a metric and quantum Levi-Civita connection on a set with 2 points with its universal differential calculus. Not too surprisingly, it has zero quantum Riemannian curvature. 
 
  \subsubsection{Quantum BV algebra} In the noncommutative case the $\delta$ in Theorem~\ref{perptheorem} no longer has to obey the 7-term identity for a BV algebra as was the case classically in Section~\ref{secrec}. But for examples close to classical it should deform that. Therefore  according to our philosophy a differential BV algebra should perhaps be understood as a quantum metric and covariant derivative giving a cocycle extension as we have seen. It could therefore be interesting to interpret and extend BRST methods in this way.


\begin{thebibliography}{}
 
 \bibitem{AmeMa:wav} G. Amelino-Camelia and S. Majid, Waves on noncommutative spacetime and gamma-ray bursts, Int. J. Mod. Phys. A 15 (2000) 4301--4323
 
 \bibitem{BV}I.A. Batalin, G.A., Vilkovisky, Quantization of gauge theories with linearly dependent generators,  Phys. Rev. D 28 (1983) 2567--2582

\bibitem{Beggs} G. Alhamzi, E.J. Beggs \&  A.D. Neate, From homotopy to It\^o calculus to Hodge theory, arXiv:1307.3119 (math.QA)

\bibitem{BegMa2}
E.J. Beggs \& S. Majid, *-Compatible connections in noncommutative Riemannian geometry, J. Geom. Phys. (2011) 95--124

\bibitem{BegMa3}
E.J. Beggs \& S. Majid, Gravity induced by quantum spacetime, in press Classical and Quantum Gravity (2014) 40pp

\bibitem{MDV}M. Dubois-Violette \& P.W. Michor, Connections on central bimodules in noncommutative differential geometry, J. Geom. Phys. 20 (1996) 218 --232

\bibitem{massivegrav} K. Hinterbichler, Theoretical aspects of massive gravity, Rev. Mod. Phys. 84 (2012)  671 

\bibitem{Ma:pla} S. Majid,  Hopf algebras for physics at the Planck scale, Classical and Quantum Gravity 5 (1988) 1587--1607

\bibitem{Ma:mom} S. Majid, Braided momentum in the $q$-Poincare group, J. Math. Phys. 34 (1993) 2045--2058

\bibitem{Ma:gra}  S. Majid, Noncommutative Riemannian geometry of graphs, J. Geom. Phys. 69 (2013) 74--93

\bibitem{Ma:book}S. Majid, Foundations of Quantum Group Theory, Cambridge Univ. Press, 1996

\bibitem{Ma:alm} S. Majid, Almost commutative Riemannian geometry: wave operators,  Commun. Math. Phys. 310 (2012) 569--609

\bibitem{Ma:icnfp1} S. Majid, Newtonian gravity on quantum spacetime,  in Proc. ICNFP 2012; in press Euro Phys. J. Web of Conf, 10pp

\bibitem{Ma:rec}S. Majid, Reconstruction and quantisation of Riemannian structures, arXiv:1307.2778 (math.QA)


\end{thebibliography}
\end{document}